\documentclass[reprint,aip,amsmath,amssymb,longbibliography,floatfix,10pt,tightenlines]{revtex4-2}

\usepackage[english]{babel}
\usepackage{bm,graphicx,braket,color,booktabs,fixmath}
\usepackage[hidelinks,colorlinks=true,citecolor=black,linkcolor=black,urlcolor=black]{hyperref}
\usepackage{algorithm}
\usepackage{array}
\usepackage[noend]{algpseudocode}
\makeatletter

\def\bbl@set@language#1{%
  \edef\languagename{%
    \ifnum\escapechar=\expandafter`\string#1\@empty
    \else\string#1\@empty\fi}%
  \@ifundefined{babel@language@alias@\languagename}{}{%
    \edef\languagename{\@nameuse{babel@language@alias@\languagename}}%
  }%
  \select@language{\languagename}%
  \expandafter\ifx\csname date\languagename\endcsname\relax\else
    \if@filesw
      \protected@write\@auxout{}{\string\select@language{\languagename}}%
      \bbl@for\bbl@tempa\BabelContentsFiles{%
        \addtocontents{\bbl@tempa}{\xstring\select@language{\languagename}}}%
      \bbl@usehooks{write}{}%
    \fi
  \fi}
\newcommand{\DeclareLanguageAlias}[2]{%
  \global\@namedef{babel@language@alias@#1}{#2}%
}
\makeatother

\DeclareLanguageAlias{en}{english}

\begin{document}

\title{Inverse design of photonic devices with strict foundry fabrication constraints}

\author{Martin F. Schubert}
\email{mfschubert@gmail.com}
\thanks{Authors appear in order of contribution.}
\author{Alfred K. C. Cheung}
\email{alfredkcc@x.team}
\author{Ian A. D. Williamson}
\email{iwill@x.team}
\author{Aleksandra Spyra}
\email{aspy@x.team}
\affiliation{X Development LLC, 100 Mayfield Ave, Mountain View, CA 94043}
\author{David~H.~Alexander}
\email{dhalexander@google.com}
\affiliation{Google, 1600 Amphitheatre Pkwy, Mountain View, CA 94043}

\date{\today}

\begin{abstract}
We introduce a new method for inverse design of nanophotonic devices which guarantees that resulting designs satisfy strict length scale constraints --- including minimum width and spacing constraints required by commercial semiconductor foundries. The method adopts several concepts from machine learning to transform the problem of topology optimization with strict length scale constraints to an unconstrained stochastic gradient optimization problem. Specifically, we introduce a conditional generator for feasible designs and adopt a straight-through estimator for backpropagation of gradients to a latent design. We demonstrate the performance and reliability of our method by designing several common integrated photonic components.
\end{abstract}

\maketitle

\section{Introduction}

Integrated photonic devices have led to game changing new capabilities in applications ranging from high-speed communications \cite{Marpaung2019-jx} to next-generation quantum computing platforms \cite{Arrazola2021-ms} and machine learning hardware accelerators \cite{Wetzstein2020-lz}.
These platforms stand to benefit from photonic components with improved performance, lower losses, larger bandwidths, and more compact footprints.
However, achieving such multi-faceted specifications is extremely challenging through conventional intuition-backed design methodologies.
In contrast, computational inverse design techniques can efficiently explore complex and high-dimensional design landscapes which are inaccessible to human designers.
Enabled by adjoint variable methods \cite{Veronis2004-zk, Molesky2018-oe}, gradient-based optimization techniques have proven to be among the most successful at producing photonic devices with high performance, with examples including wavelength-selective elements (e.g. multiplexers, filters, and resonators), signal routing elements (e.g. waveguide bends and couplers), and active components (e.g. modulators) \cite{Sell2017-hs, Hughes2018-rj, Jin2018-vz, Wang2018-wi, Lin2019-dt, Piggott2020-oa, Yang2020-jd, Sapra2020-zo, Minkov2020-cu, Tseng2021-jh, Hammond2021-pv}.

The approach in gradient-based design strategies is to first formulate an objective function, calculated in terms of a device's simulated performance.
Adjoint methods allow these objective functions and their gradients to be efficiently computed at a cost of only two full-wave simulations \cite{Veronis2004-zk, Molesky2018-oe}, independently of the number of degrees of freedom in a design.
When used in the context of modern automatic differentiation frameworks (e.g. JAX and TensorFlow), these techniques enable flexible end-to-end device optimization frameworks \cite{Minkov2020-cu, Hughes2019-qu}.
However, guaranteeing that an optimized design yields the desired performance, while simultaneously satisfying the numerous design rules of modern semiconductor manufacturing processes \cite{Liebmann2001-ta, Orcutt2010-sl}, still poses a major challenge.
Particularly relevant for integrated photonic devices are the capabilities of the lithographic systems, which define constraints such as the minimum feature length scale that can be reliably printed.
Elements of a design which are smaller than this length scale may print inconsistently or may be completely absent in the fabricated structure.
Moreover, inclusion of a design feature with sub-resolution size can have broad yield implications.
Thus, semiconductor foundries specify design rules which define the minimum width and minimum spacing of design features.
Compliance with these rules, as validated by a design rule check (DRC), is typically a prerequisite for fabrication.

There have been many proposals for DRC-compliant inverse design strategies.
One class of strategies selects a design parameterization with intrinsic geometric length scale guarantees, e.g. by optimizing the placement of geometric primitives which already satisfy the desired minimum feature size constraints \cite{Wang2020-cc}.
A related approach uses a parameterization which enables analytic constraints to be applied directly to the design degrees of freedom, e.g. by optimizing the coordinates of polygon vertices comprising shapes in the design mask \cite{Michaels2018-vd}.
However, the drawback in both of these strategies is that they offer up a limited landscape of potential designs.
For example, when optimizing the vertices of a polygon, an optimizer will be unable to alter the topology of a design by closing or opening holes.

In contrast, topology optimization strategies open up a much larger design landscape by parameterizing a design as a pixelated image, which is typically transformed by a sequence of convolutional filters and pixel-wise nonlinear functions \cite{Frei2007-pl, Jensen2011-br, Wang2012-iv, Qian2013-cc, Zhou2015-hg, Hagg2018-ku, Hammond2021-pv}.
These so-called density or three-field parameterizations allow a gradient-based optimization algorithm (e.g. LBFGS or MMA \cite{Svanberg1987-ya}) to modify any pixel within the design and apply arbitrary changes to the topology.
In practice however, there is often a trade-off between maintaining design feasibility (binarization of the pixels and feature size constraints) and attaining high performance.
This is because a continuous optimizer must traverse \textit{infeasible} regions of the design landscape before reaching \textit{feasible} regions, which may or may not have high performance.
The most common strategy in topology optimization is to begin with a continuous optimization phase before gradually ramping up binarization and applying design rule constraints.
However, if binarization is ramped up too quickly, a continuous optimizer will stall and fail to make progress towards improving performance.
Similarly, constraints or penalties on the feature size must be introduced in a way which will not push the optimizer away from high performance regions of the parameter space.
Often in these strategies, the selection of optimization hyper parameters and the detailed strategy for enforcing design feasibility is an art and, moreover, depends on both the type of device being optimized as well as the performance objective.

In this work we propose an \textit{always feasible} gradient-based inverse design framework with strict guarantees on DRC-compliance.
The key features of our proposal are a \emph{conditional generator} for feasible designs, which is combined with a \emph{straight-through estimator} (STE) to enable gradient backpropagation. 
The combination of these two components yields a differentiable transform which can be incorporated into an computational graph, much like a convolutional filter or projection operator in density-based topology optimization.
We apply our framework to several practical photonic design challenges (a waveguide bend, a beam splitter, a waveguide mode converter, and a demultiplexer), and demonstrate that the approach reliably yields designs with high performance.

\section{Methodology}

\subsection{Brush-based minimum length scale constraints}

In this section, we introduce a mathematical framework for measuring whether a binarized pixel array is DRC-compliant, focusing on the minimum width and minimum spacing rules commonly required by semiconductor foundries.
In our framework, we consider binary arrays with values of $-1$ and $+1$, referring to these as \textit{void} and \textit{solid}, respectively.
During the course of a fabrication workflow, such designs may be converted to contours by off-the-shelf algorithms, such as marching squares or dual contouring \cite{Ju2002-jj}.

\begin{figure}
    \centering
    \includegraphics[width=3.3in]{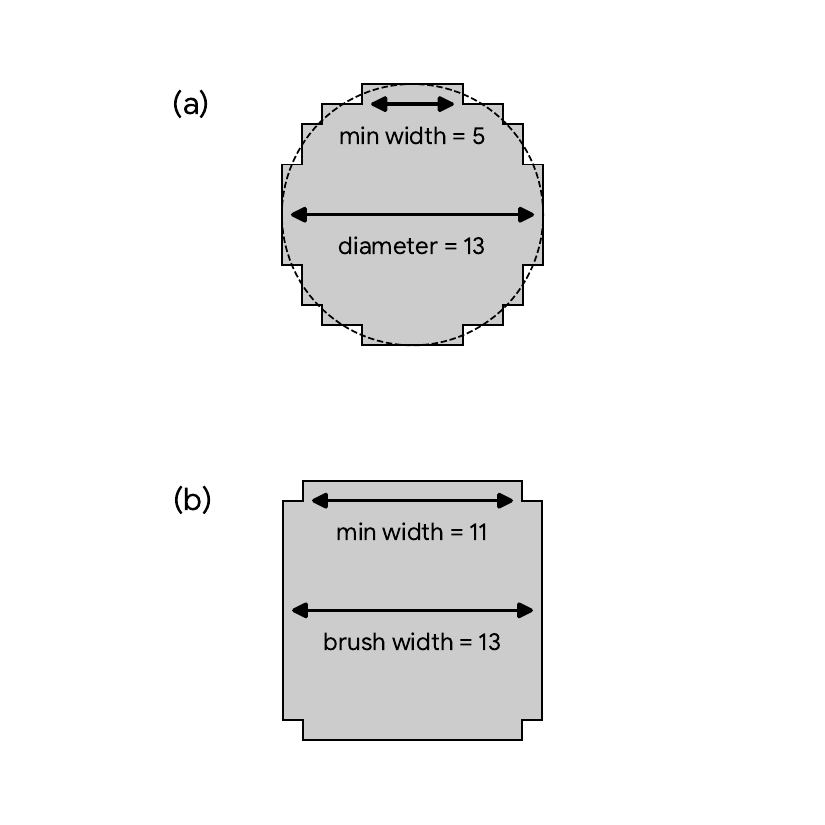}
    \caption{(a) Circular brush with diameter 13, and (b) notched square brush with width 13.}
    \label{fig:brushes}
\end{figure}

All binary arrays can be realized by sequentially setting individual pixels to void or solid; conceptually, this is equivalent to the repeated placements and translations of single-pixel void or solid brushes. While a single-pixel brush can generate all possible binary arrays, some arrays are infeasible with larger brushes. In the trivial case, a single-pixel solid feature cannot be realized with a $3\times3$ pixel solid brush, since centering the brush on a single pixel also sets adjacent pixels to solid. Throughout this paper, we refer to the pixel on which the brush is centered as a touch, $t$ of the brush.

Our goal is to determine whether a given design, $x$ is feasible with a some brush, $b$, i.e. to determine whether all its solid and void features can be realized with $b$. When this is true, the condition
\begin{equation}
    \mathcal{O}(x, b) = \lnot \mathcal{O}(\lnot x, b) = x
    \label{eqn:condition}
\end{equation}
holds, where $\mathcal{O}(x, b)$ is the morphological opening of $x$ with brush $b$ and $\lnot$ is a logical \textit{not} operation which changes solid features into void features and void features into solid features, i.e. a value of -1 to 1 and a value of 1 to -1. The morphological opening operation is computed by the application of a binary erosion, $\mathcal{E}$ followed by a dilation \cite{erosion_dilation}, $\mathcal{D}$ with the same brush, i.e. $\mathcal{O}(x, b) = \mathcal{D}(\mathcal{E}(x, b), b)$. Intuitively, the dilation operation causes features in an image to expand while erosion causes features in an image to shrink.
An illustration of these operations is provided in the supporting information.
For a $b$-feasible design $x$, $\mathcal{E}(x, b)$ erodes the solid features and recovers the solid brush touches, while $\mathcal{E}(\lnot x, b)$ recovers the void brush touches. Dilation of the touches then restores the solid and void pixels, respectively. Intuitively, if $x$ contains a small feature that is incompatible with $b$, then erosion will remove it entirely and no solid pixels will be restored by the dilation. Thus, the binary opening operation removes small features, while leaving feasible features unchanged.

\begin{figure}
    \centering
    \includegraphics[width=3.3in]{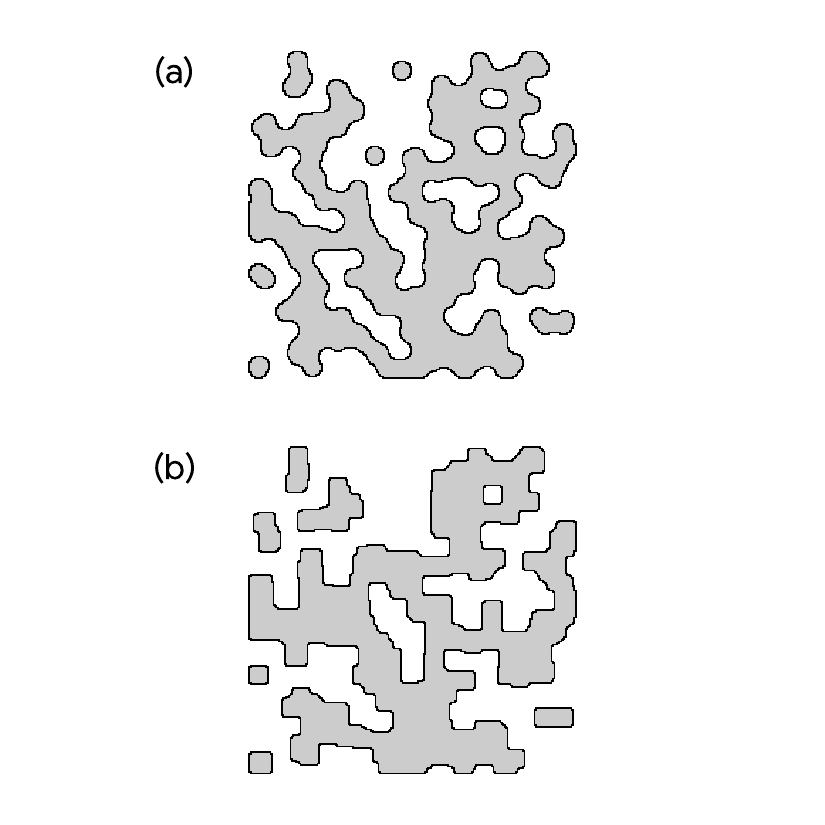}
    \caption{Random designs feasible for (a) diameter-13 circular brush, and (b) width-13 notched square brush.}
    \label{fig:designs}
\end{figure}

Length scale in a topology optimization context is typically associated with the minimum diameter of circular features. Thus, we say that if a design can be realized with an approximately circular brush $b_L$ with a diameter of $L$ pixels, it has a minimum length scale of at least $L \cdot d$, where $d$ is the design pixel pitch. The minimum length scale of the design is the largest value of $L$ for which Eqn. (\ref{eqn:condition}) is satisfied with brush $b_L$. When contours are obtained by dual contouring (i.e. outlining pixels, for binary designs), a design with minimum length scale will also have related minimum width and spacing, equal to the minimum width of the brush itself. An example circular brush with associated measurements is shown in Fig. \ref{fig:brushes}a and an example design which is feasible with the circular brush is shown in Fig. \ref{fig:designs}a.

We turn now to minimum width and spacing constraints which result from foundry design rules. Given a minimum width and spacing constraint \cite{koefferlein, calibre}, there are many brushes for which the associated feasible designs satisfy the constraint. The smallest of these is the notched square, i.e. a square brush which is solid everywhere except at the corners, as shown in Fig. \ref{fig:brushes}b. An example design is shown in Fig. \ref{fig:designs}b. When the constraint requires a physical minimum feature width and spacing of $w$, designs feasible with the notched square brush of width $L = w / d + 2$ will satisfy the constraint. Thus, if we can require that our designs are feasible with a notched square brush of width $L = w  / d + 2$, we are also guaranteed to satisfy width and spacing rules of size $w$.

Although reasoning about minimum width and spacing is straightforward for dual-contoured designs feasible with the notched-square brush, we have also extensively validated that such designs are design-rule compliant using industry standard design software such as KLayout \cite{koefferlein} and Siemens Calibre Physical Verification \cite{calibre}.

\subsection{Generator for feasible designs}

In this section we describe our method of inverse design subject to strict length scale constraints. We have shown that designs which are feasible with a certain brush $b$ ($b$-feasible) will satisfy minimum length scale or width and spacing constraints. Properly, this can be considered a combinatorial optimization problem. However, our approach is to approximately solve this problem, by transforming it into an unconstrained stochastic gradient optimization problem. Critically, we use a conditional generator for feasible designs and a straight-through estimator (STE) for backpropagation; these two components are discussed subsequently.

\begin{figure}[t!]
    \centering
    \includegraphics[width=3.3in]{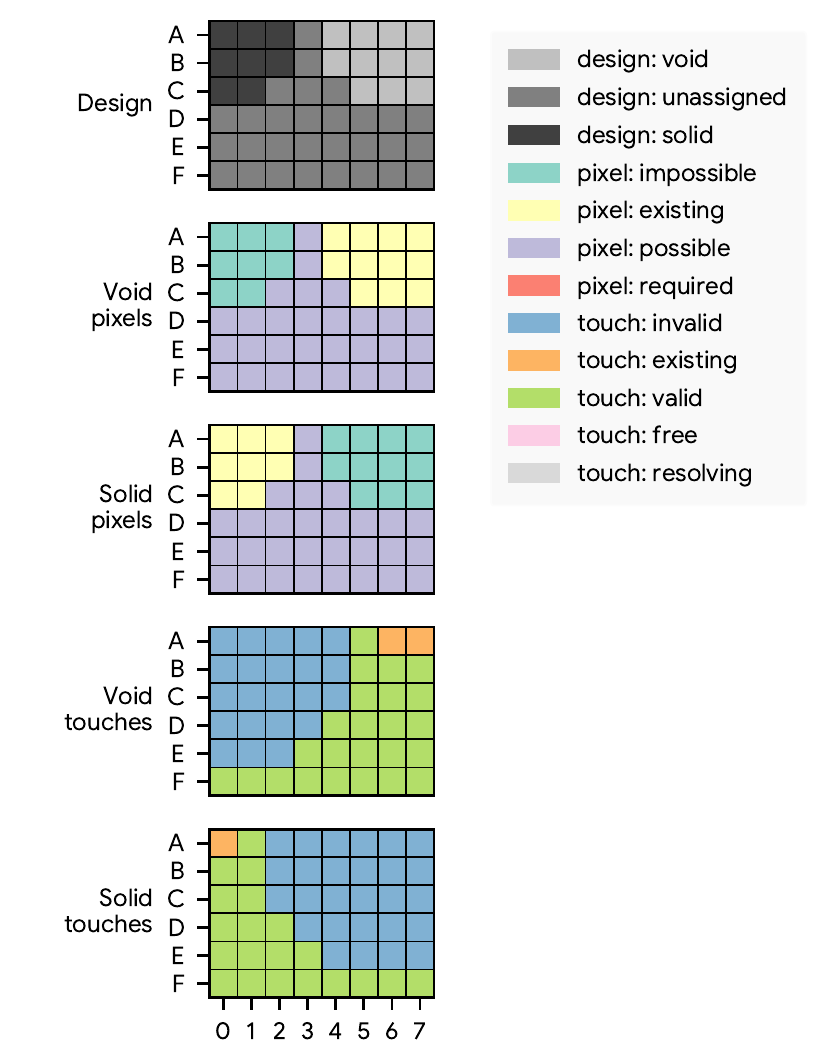}
    \caption{Design, pixels, and touches for an example traversal using the width-5 notched square brush.}
    \label{fig:traversal_snapshot}
\end{figure}

In general, a binary design $x$ can be produced by dilation with $b$ of the solid and void touches $t_b^s$ and $t_b^v$, i.e.
\begin{equation}
x = \mathcal{D}(t_b^s, b) =\lnot \mathcal{D}(t_b^v, b)
\label{eqn:feasible}
\end{equation}
We consider the generation of $b$-feasible arrays as a process in which $t_b^s$ or $t_b^v$ are updated with a single-element change at each step. In this context, we introduce several touch states and pixel states which may be associated with each location at any step in the process. Defining these states for the solid case: \emph{existing} solid pixels are those which are assigned to solid by some touch in $t_b^s$,
\begin{equation}
p_{\text{existing}}^s = \mathcal{D}(t_b^s, b).
\end{equation}
\emph{Impossible} solid touches are those which would assign an existing void pixel to solid, and consists of all pixels within some distance of existing void pixels,
\begin{equation}
t_{\text{impossible}}^s = \mathcal{D}(p_{\text{existing}}^v, b).
\end{equation}
\emph{Valid} solid touches are those which are not impossible and not already in $t_b^s$,
\begin{equation}
t_{\text{valid}}^s = \lnot t_{\text{impossible}}^s \land \lnot t_b^s,
\end{equation}
where $\land$ is an element-wise logical \textit{and} operation. \emph{Possible} solid pixels are those obtained by dilating all existing or valid touches with $b$, i.e. they are those pixels which are or may be solid in a feasible array,
\begin{equation}
p_{\text{possible}}^s = \mathcal{D}(t_b^s \lor t_{\text{valid}}^s, b),
\end{equation}
where $\lor$ is an element-wise logical \textit{or} operation. \emph{Required} solid pixels are those which are not existing and not possible for void,
\begin{equation}
p_{\text{required}}^s = \lnot p_{\text{existing}}^s \land \lnot p_{\text{possible}}^v.
\end{equation}
\emph{Resolving} solid touches are those which are valid and would assign solid to a required-solid pixel,
\begin{equation}
t_{\text{resolving}}^s = \mathcal{D}(p_{\text{required}}^s, b) \land t_{\text{valid}}^s.
\end{equation}
Finally, \emph{free} touches are valid but only color existing pixels.
\begin{equation}
t_{\text{free}}^s = \lnot \mathcal{D}(p_{\text{possible}}^v \lor p_{\text{existing}}^v, b) \land t_{\text{valid}}^s. 
\end{equation}
Corresponding void touches and pixels are be computed using the process. Fig. \ref{fig:traversal_snapshot} shows an example of pixels and touches computed from the steps above and a full example is illustrated in the Supporting Information.

A scheme that generates feasible arrays is outlined in Algorithm \ref{algorithm:generator}. The algorithm begins with empty $t_b^s$ and $t_b^v$ and loops until all pixels are existing for either solid or void. At each step, the pixel and touch states are updated. If free touches exist, all are selected. If resolving touches exist, one of these is selected; otherwise, any of the valid touches are selected.

\begin{figure}[t!]
\begin{algorithm}[H]
\caption{Generator for $b$-feasible designs}
\begin{algorithmic}
\Procedure{Generator}{$b$}
    \State initialize empty $t_b^s$ and $t_b^v$
    \While{design is incomplete}
        \State update pixel and touch states
        \If{free touches exist}
            \State select all free touches
        \ElsIf{resolving touches exist}
            \State select a single resolving touch
        \Else
            \State select a single valid touch
        \EndIf
    \EndWhile
\EndProcedure
\end{algorithmic}
\label{algorithm:generator}
\end{algorithm}
\end{figure}

The algorithm sets at least one pixel at every step, and thus requires at most $N$ iterations for a reward array with $N$ pixels. A basic implementation of the algorithm involves full-array erosion and dilation operations at each step, which have cost proportional to $O(N)$, and leading to an overall cost $O(N^2)$. However, it is straightforward to implement windowed operations which have fixed cost independent of $N$, and so feasible design generation with $O(N)$ complexity is possible.

If touch selection is random, the algorithm simply yields a random feasible design. In practice, we greedily select the best touch as computed from a pixel reward array $\theta$, where $\theta$ is obtained from a latent design produced in the course of an optimization, and is fixed during each run of the generator. The total reward of a solid touch is the sum of $\theta$ elements that would be set to solid were the touch to be made, i.e. the sum of $\theta$ elements set by the touch. The reward of a void touch is the negative sum of elements set to void by the touch. A simple illustrated example of how rewards for touches are calculated can be found in the supporting information. With this change, the algorithm becomes a conditional generator of feasible designs, where $\theta$ biases the feasible design generation. The designs in Fig. \ref{fig:designs} have been created by running the algorithm on the same random reward array.

To gain some intuition for why this scheme always produces feasible designs, consider that the algorithm will select exclusively solid or exclusively void until reaching a state where all remaining pixels are possible for both solid and void. In this state, no touch can lead to an invalid status. Since the starting state is one where all pixels are possible for both solid and void, we can see that the algorithm will never produce an invalid status.

We emphasize that the full space of feasible designs of brush $b$ is accessible by Algorithm \ref{algorithm:generator}. This can be clearly seen, since e.g. if $\theta = x$ is provided, where $x$ is any feasible design, the output of the algorithm will simply be $x$.

\subsection{Straight through estimator}
Algorithm \ref{algorithm:generator} is not differentiable, and so it cannot be directly used with backpropagation in a gradient-based optimization setting. Drawing inspiration from the field of binary and quantized neural networks, we note that binary optimization can be viewed as topology optimization with a minimum length scale equal to a single pixel. Binary and quantized neural networks often make use of a STE in training, to backpropogate gradients through binary activation functions or to latent weights from which low-precision weights are obtained \cite{bengio2013estimating}. Specifically, in this approach one substitutes the gradient of a non-differentiable function with that of an estimator, i.e.
\begin{equation}
    \frac{\partial f}{\partial x} \approx \frac{\partial}{\partial x}f_\text{estimator}
\end{equation}
A typical estimator used for binarization (e.g. via application of the sign function) is the identity operation.

\subsection{Computational graph}

\begin{figure}
    \centering
    \includegraphics[width=3.3in]{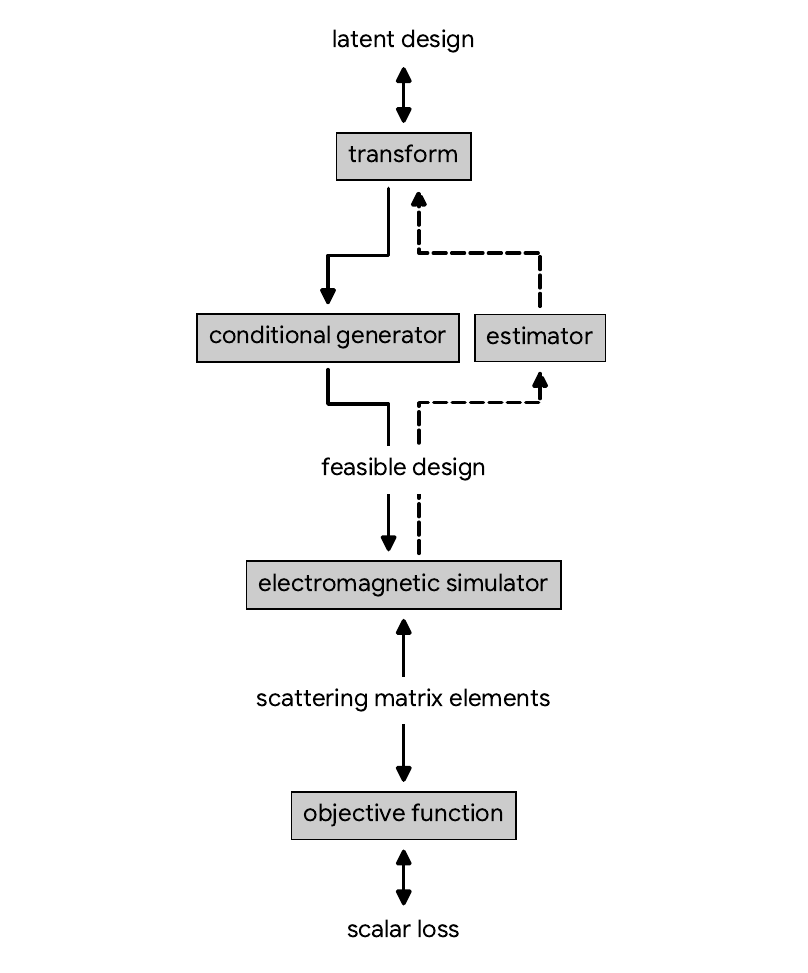}
    \caption{Computational graph of our inverse design scheme. Downward arrows indicate the forward computation, while upward arrows correspond to backpropagation. Gray boxes represent operations and text represents arrays or results.}
    \label{fig:graph}
\end{figure}

The complete computational graph for the inverse design problem is depicted in Fig. \ref{fig:graph}. A latent design is passed through a transform, yielding the reward array $\theta$. This is passed to the conditional generator, which produces a feasible design. We pass the design to an electromagnetic simulation engine which computes scattering matrix elements; these are the inputs to an objective function, which computes a scalar loss value. Optionally, we symmetrize the transform output to favor symmetric feasible designs. All operations in this graph are differentiable, with the exception of the generator --- for which we use the STE. Thus, we can compute an estimated gradient of the loss with respect to the latent design.

For the transform, we have found that
\begin{equation}
\label{eqn:estimator}
    D(t, b, \beta) = \tanh(\beta (t \circledast b))
\end{equation}
is a good choice, where $\beta$ is a scalar hyperparameter in the range 2-8. The convolutional transform aids in the optimization, as it ensures that the reward array $\theta$ is smooth, restricted in range, and has an associated length scale, determined by the brush size.

The form of our estimator matches that of the transform. As with the transform, outputs of the estimator have a characteristic length scale and are restricted to the range $(-1, 1)$, just as outputs of the generator itself are. Thus, the estimator can be seen as a differentiable approximation of the conditional generator. The success of this estimator is consistent with the finding in binary neural networks that estimators which approximate their forward-pass counterpart outperform simpler functions \cite{yin2019understanding}.

The latent design is randomly initialized with a bias so that the first feasible design is fully solid, and it is updated iteratively using the Adam scheme \cite{kingma2017adam}, which is a common choice for stochastic optimization problems in machine learning. We use a learning rate of 0.01, and $\beta_1=0.667$ and $\beta_2=0.9$ for the decay of the gradient and second moment exponential moving averages, respectively.

Notably, our method does not require scheduling of hyperparameters or changes in parameterization during the course of an optimization, and gives a feasible design at every optimization step. We expect that this simplicity could be advantageous from the perspective of a photonic designer tasked with creating a new component.

\subsection{Objective function}

In a practical design application, one is generally concerned with creating a component that satisfies a performance specification, i.e. a desired level of insertion loss, return loss, and cross talk. We define such specification by defining a cutoff value $|S_{\text{cutoff}}|$ for the magnitude for each scattering parameter element of $S$. Additionally, there is an implicit bound on the allowed values, which is imposed by physical constraints. For example, transmission magnitudes greater than unity are not possible for passive photonic components.
Our scalar objective function has the form
\begin{equation}
    \mathcal{L}(S) = \left(\left\vert\left\vert\text{softplus}{\left(g \frac{\vert S\vert^2 -\vert S_{\text{cutoff}}\vert^2}{\min(w_{\text{valid}})}\right)}\right\vert\right\vert_2\right)^2,
    \label{eqn:loss_fn}
\end{equation}
where $g$ is a vector of signs (+1 or -1) matching the size of $S$.
The values of $g$ are positive (+1) where the corresponding element of the cutoff array, $|S_{\text{cutoff}}|$ specifies the maximum value, and negative (-1) where $|S_{\text{cutoff}}|$ specifies the minimum value.
$w_{\text{valid}}$ also has the same size as $S$ and defines the range of allowed values for a scattering parameter element, which is defined as the distance between the performance cutoff, specified by $|S_{\text{cutoff}}|$, and the physical bound on the scattering parameter element.
For example, a minimum transmission amplitude cutoff of $0.5$ (-3 dB in power transmission) would have a $w_{\text{valid}}$ of $0.5$, because the maximum transmission possible is 1.0.
All operations other than the min in the softplus of Eqn. (\ref{eqn:loss_fn}) are element-wise.

\section{Results}

\subsection{Nanophotonic optimization problems}

In this section we apply our proposed inverse design framework to several integrated photonic components operating in the O-band, demonstrating designs for a waveguide bend, a mode converter, a beamsplitter, and a wavelength demultiplexer \cite{ceviche_challenges}.
All components are optimized for their characteristics within two 10 nm wavelength bands centered at 1270 nm and 1290 nm.
Performance targets for the components are specified in terms of their scattering parameters, and are summarized in Table \ref{tab:specs}.
Simulations of the components are performed using Ceviche \cite{Hughes2019-qu}, an open source 2D finite difference frequency domain (FDFD) simulator. 
Silicon ($\varepsilon_r = 12.25$) and silicon oxide ($\varepsilon_r = 2.25$) are used for solid and void materials, respectively.
All components are coupled to waveguides having 400 nm width, and we deal with the fundamental waveguide mode unless noted otherwise.

\begin{table*}
\begin{tabular}{| r  r | c | c | c | c |}
\hline
&  
& \begin{tabular}{@{}c@{}}Waveguide \\ bend\end{tabular} 
& \begin{tabular}{@{}c@{}}Mode \\ converter\end{tabular} 
& \begin{tabular}{@{}c@{}} \ \\ Beamsplitter \end{tabular} 
& \begin{tabular}{@{}c@{}}Wavelength \\ demultiplexer\end{tabular}  \\
\hline
 $S_{11}$& 1270 nm & $\leq$ -20 dB  & $\leq$ -20 dB  & $\leq$ -20 dB  & $\leq$ -20 dB \\
         & 1290 nm & $\leq$ -20 dB  & $\leq$ -20 dB  & $\leq$ -20 dB  & $\leq$ -20 dB \\
         \hline
 $S_{21}$& 1270 nm & $\geq$ -0.5 dB & $\geq$ -0.5 dB & $\geq$ -3.5 dB & $\geq$ -3 dB  \\
         & 1290 nm & $\geq$ -0.5 dB & $\geq$ -0.5 dB & $\geq$ -3.5 dB & $\leq$ -20 dB \\
         \hline
 $S_{31}$& 1270 nm & -              & -        & $\geq$ -3.5 dB & $\leq$ -20 dB \\
         & 1290 nm & -              & -        & $\geq$ -3.5 dB & $\geq$ -3 dB  \\
         \hline
 $S_{41}$& 1270 nm & -              & -            & $\leq$ -20 dB  & -           \\
         & 1290 nm & -              & -            & $\leq$ -20 dB  & -           \\
\hline
\end{tabular}
\caption{Performance specification for component power scattering parameters at the 1270 nm and 1290 nm wavelength bands.}
\label{tab:specs}
\end{table*}

The waveguide bend features a $1.6\times1.6 \ \mu\text{m}^2$ design region, with waveguides connecting to the left (port 1) and bottom (port 2). Given excitation from port 1, we aim to maximize transmission to port 2 while keeping backreflection low. Diagonal reflection symmetry is imposed, so that excitation from port 2 yields identical behavior.

The spatial mode converter features a $1.6\times1.6 \ \mu\text{m}^2$ design region, with waveguides connecting to the left and right. We seek designs that maximally convert the fundamental waveguide mode on the left (port 1) to the second order mode on the right (port 2), with minimal backreflection into port 1.

The beamsplitter design region is $3.2\times2.0 \ \mu\text{m}^2$ in size, with two waveguides connecting to the left (port 1 and port 4) and two waveguides connecting to the right (port 2 and port 3). Given excitation from port 1, the beamsplitter aims to divide power equally into port 2 and port 3 while minimizing backreflection into port 1 or transmission into port 4. We impose reflection symmetry along the horizontal and vertical axes, so that excitation from any other port yields identical behavior.

The wavelength demultiplexer features a $6.4\times6.4 \ \mu\text{m}^2$ design region, with one waveguide connecting to the left (port 1) and two waveguides connecting to the right (port 2 and port 3). Given excitation from port 1, wavelengths in the first band are directed to port 2 while wavelengths in the second band are directed to port 3.

In the optimization context, we consider three wavelengths per band --- the center and extremal wavelengths. The specifications are considered to be fulfilled if the criteria in \ref{tab:specs} are satisfied for all wavelengths. Our design resolution matches the 10 nm simulation resolution, so that e.g. a circular brush with diameter 10 corresponds to a 100 nm length scale.

\subsection{Designs using 100 nm circular brush}

\begin{figure}
    \centering
    \includegraphics[width=3.3in]{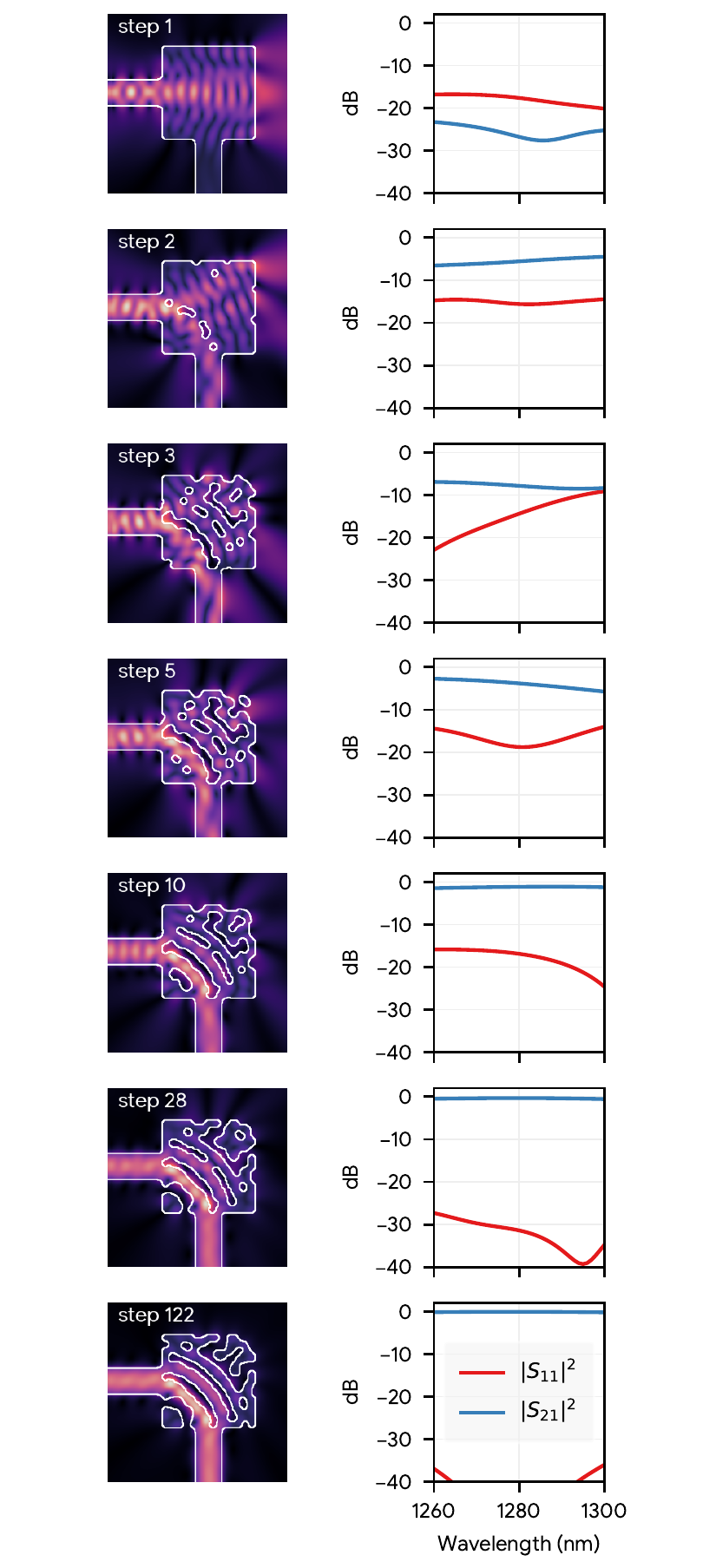}
    \caption{Evolution of a waveguide bend with 100 nm circular brush. Left: field magnitude for 1280 nm excitation with overlaid design. Right: power transmission into port 1 and 2, given excitation from port 1.}
    \label{fig:bend}
\end{figure}

Fig. \ref{fig:bend} shows optimization results for the waveguide bend with 100 nm circular brush. Upon initialization, the design is entirely silicon with no enclosed oxide. The initial design is poor, with low transmission to the output port and substantial reflection back to the input port. Subsequent updates significantly modify the topology of the design, adding enclosed oxide features and developing isolated silicon features. We emphasize the key characteristic of our scheme --- the design at every step is fully binary and satisfies the 100 nm length scale constraint.

The target performance specification is achieved in 28 optimization steps, and the design continues to improve with more iterations. The lowest loss in the first 160 steps is found at step 122. Notably, the topology of this design differs from that at step 28, due to the fusing of two void features and the elimination of a third, along with changes in the shape of some solid features. The ability of our scheme to change topology while remaining feasible stands in contrast to some other methods, where the optimizer is locked in to a fixed topology once the optimization reaches an advanced state.

\begin{figure}
    \centering
    \includegraphics{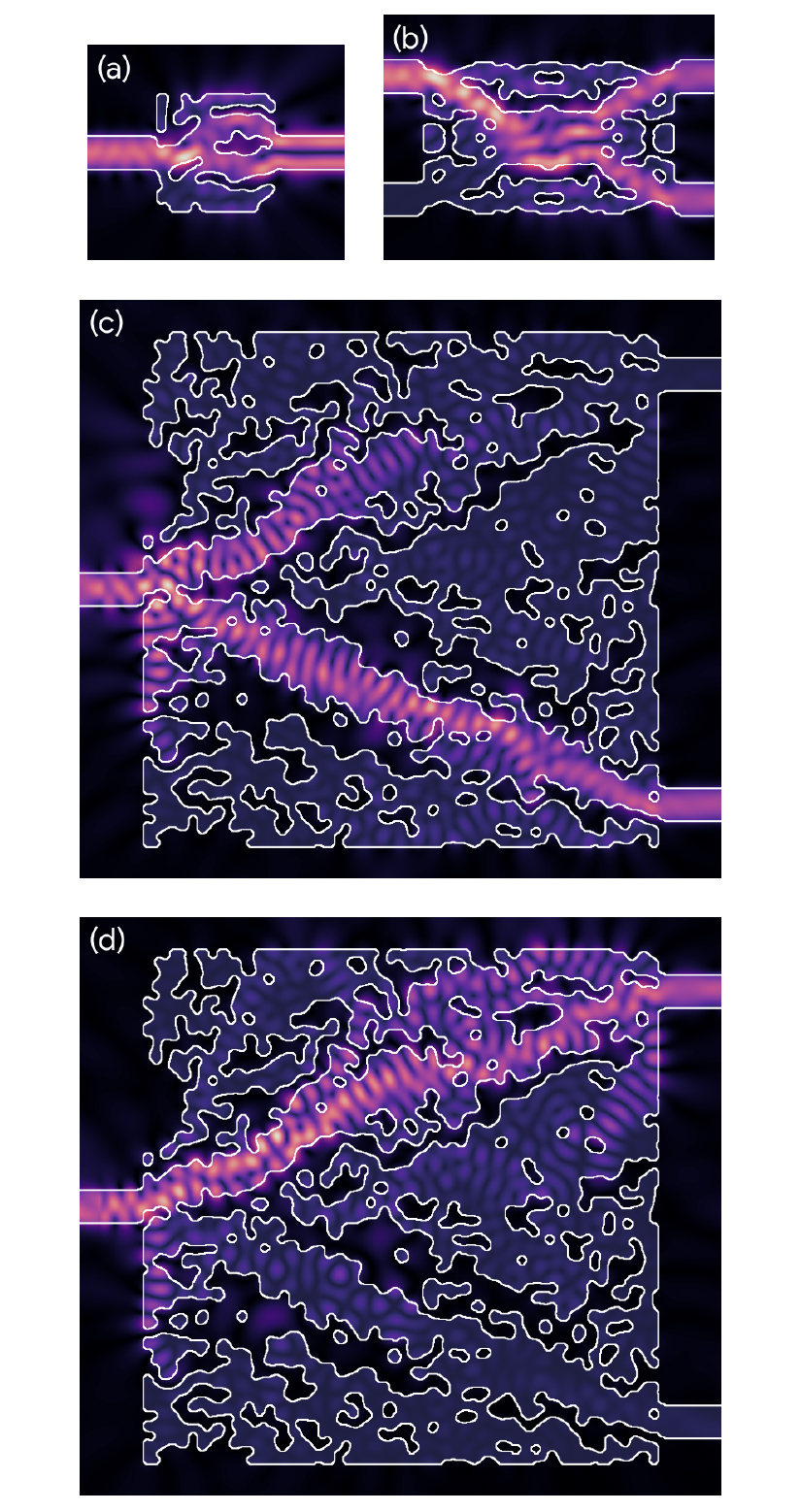}
    \caption{Lowest loss designs achieved in the first 160 steps and their respective field magnitudes for (a) mode converter and (b) beamsplitter, both with 1280 nm excitation. Design and fields for the wavelength demultiplexer, with (c) 1270 nm and (d) 1290 nm excitation. All use a 100 nm circular brush.}
    \label{fig:other_designs}
\end{figure}

\begin{figure}
    \centering
    \includegraphics{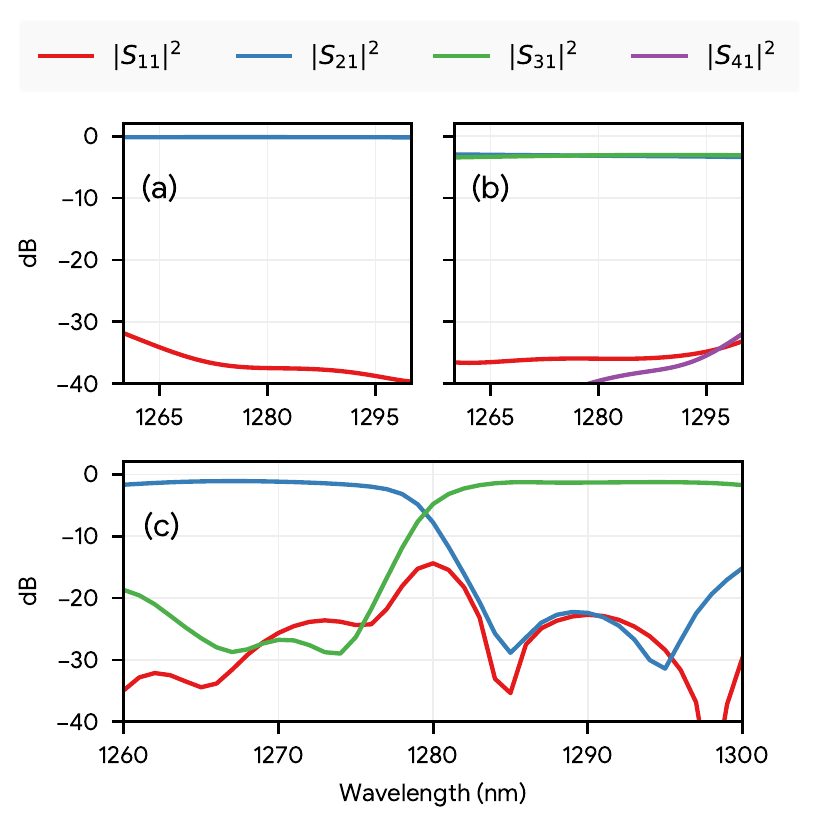}
    \caption{Scattering spectra for the (a) mode converter, (b) beamsplitter, and (c) wavelength demultiplexer with 100 nm circular brush.}
    \label{fig:other_designs_transmission}
\end{figure}

For the mode converter, beamsplitter, and wavelength demultiplexer, the lowest-loss design from the first 160 optimization steps and corresponding field magnitudes are shown in Fig. \ref{fig:other_designs}. The scattering spectra for each is shown in Fig. \ref{fig:other_designs_transmission}, where each satisfies the performance target laid out in Table \ref{tab:specs} and the 100 nm length scale constraint.

\begin{figure}
    \centering
    \includegraphics{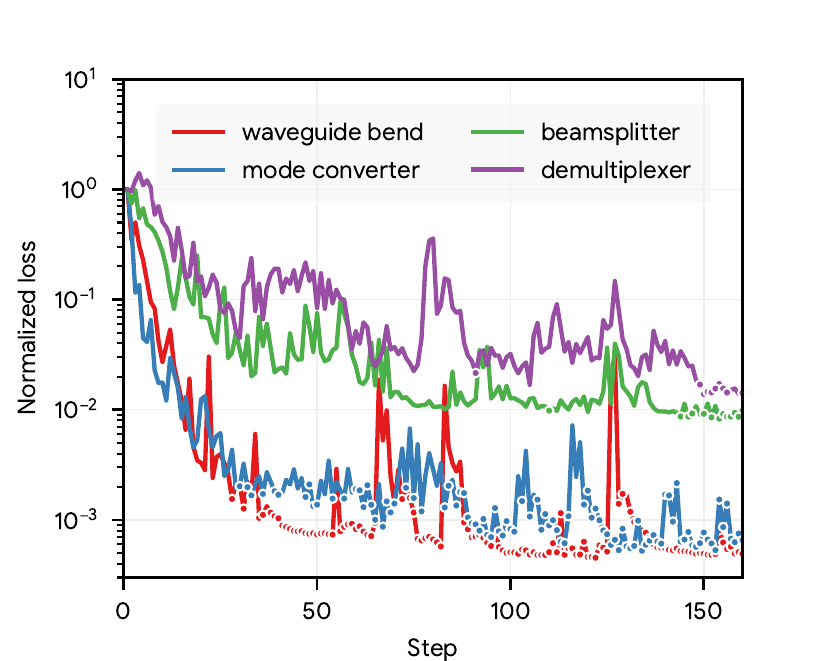}
    \caption{Normalized loss trajectories with 100 nm circular brush. Markers indicate steps where the performance target from Table \ref{tab:specs} is achieved.}
    \label{fig:loss_trajectory}
\end{figure}

As we have shown, the topology of a design can change substantially during the course of the optimization. Since the response of a device is generally discontinuous with major changes in the design shape or topology, a noisy loss trajectory is expected. This is illustrated in Fig. \ref{fig:loss_trajectory}, which shows the normalized loss value versus iteration for the waveguide bend of Fig. \ref{fig:bend} and the other three components. Points where the performance target is achieved are also highlighted. For each component, the loss generally decreases with step, but noise in the loss value and occasional regressions to higher loss can be seen.

The waveguide bend and mode converter require fewer than 40 optimization iterations to achieve the target. Meanwhile, the beamsplitter and demultiplexer appear to be relatively more difficult problems, requiring approximately 100 steps to first reach the target. However, this is still well within the range of optimization steps required in an inverse design scheme \cite{Hammond2021-pv}. In general, difficulty is directly related to the target specifications, length scale constraints, the physical size of the component, and the configuration of connecting waveguides. Reducing the length scale or increasing the design size will expand the solution space, and generally allow satisfactory solutions to be found with fewer optimization steps.

\subsection{Designs using 100 nm notched square brush}

\begin{figure}
    \centering
    \includegraphics{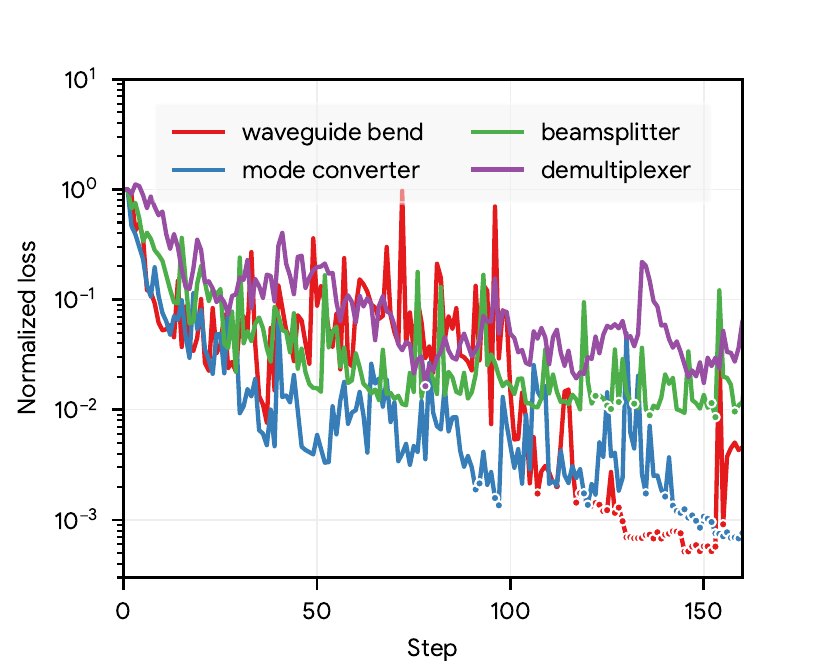}
    \caption{Normalized loss trajectories with 100 nm notched square brush. Markers indicate steps where the performance target from Table \ref{tab:specs} is achieved.}
    \label{fig:loss_trajectory_notched_square}
\end{figure}

Next, we turn to component designs generated using a 100 nm notched square brush. With the design resolution of 10 nm, these designs strictly satisfy an 80 nm minimum width and spacing constraint. The normalized loss trajectories four the components are plotted in Fig. \ref{fig:loss_trajectory_notched_square}. Scattering spectra and designs are found in the Supporting Information.

For all components, designs achieving the performance target from Table \ref{tab:specs} are found. However, the loss trajectories differ in several respects. Specifically, the trajectories are noisier, they generally contain fewer points that achieve the spec, and the spec is achieved later in the optimization. We attribute this to the fact that the notched square is relatively larger in area, limiting the design space accessible to the optimizer and making the problem more challenging. We expect that optimal hyperparameters of the Adam algorithm used to drive our optimization are somewhat problem-dependent, and in future work it would be interesting to explore various configurations. In particular, reducing the learning rate as the optimization progresses, which has been shown to aid in training of quantized neural networks \cite{10.5555/3122009.3242044}, could be beneficial.

\subsection{Reliability and effect of length scale}

Our results indicate that our inverse design framework is capable of finding designs with desired performance for a variety of fabrication-constrained photonic inverse design problems. A useful method will find such solutions for a range of length scale constraints, and do so reliably. To evaluate this, we consider the optimization problems above for circular and notched-square brushes with 60 nm, 80 nm, 100 nm, and 130 nm size. To study reliability of the method, we run 20 independent optimizations with different random initialization, each for 500 steps. Thus, for a single problem configuration (combination of device type, brush shape, and brush size), 10,000 feasible designs were generated.

\begin{figure}
    \centering
    \includegraphics{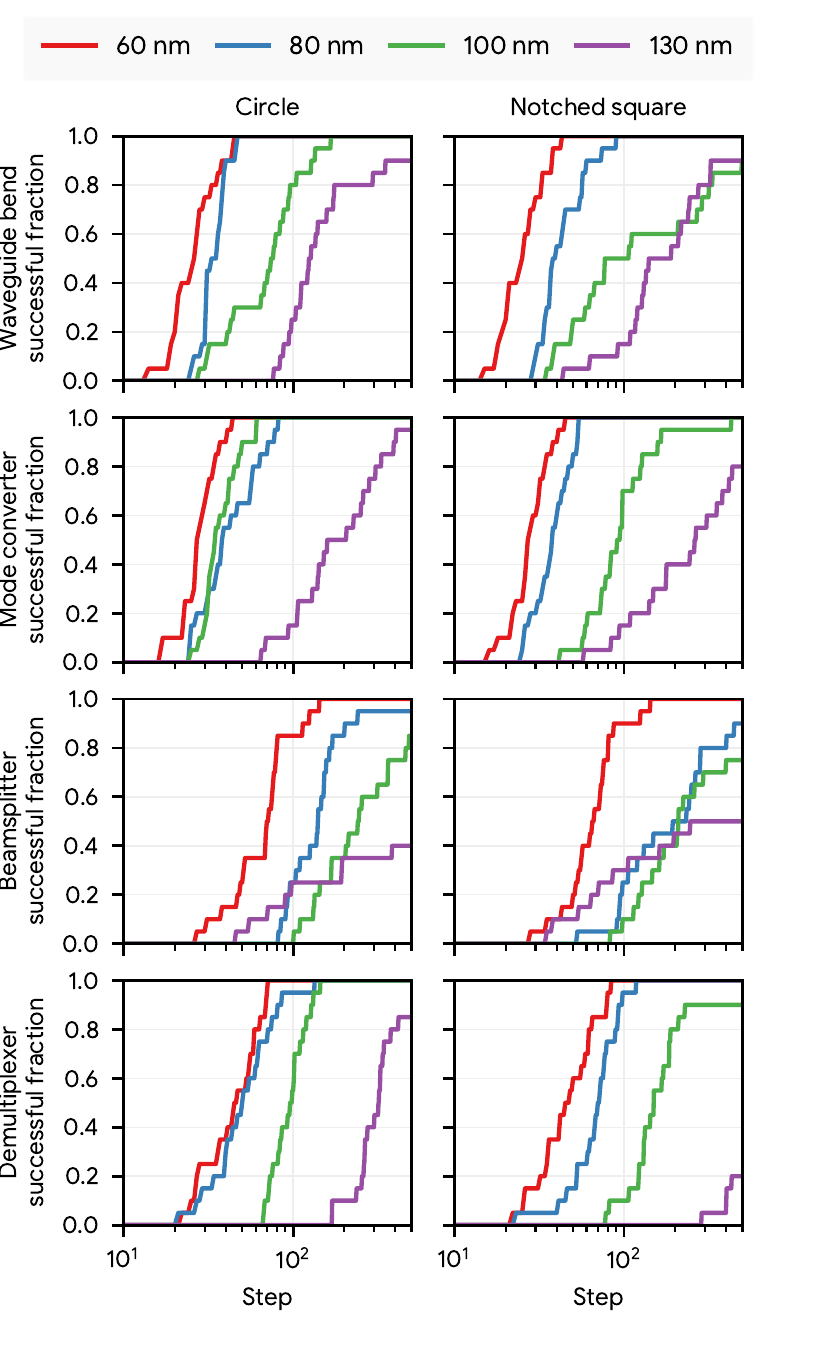}
    \caption{Fraction of independent randomly-initialized optimization runs achieving the Table \ref{tab:specs} target performance as a function of the optimization step, for each component. Left column: with circular brush. Right column: with notched square brush.}
    \label{fig:fraction_target}
\end{figure}

Our analysis proceeds as follows: for each step $i$ of an optimization run, we identify whether the target performance from Table \ref{tab:specs} is achieved at any step $j \leq i$. If the target performance is achieved at any such step, the run is considered successful at step $i$. Then, for each combination of component, brush shape, and brush size, we compute the number of runs which are successful at each step. From this, the fraction of successful runs is obtained and is shown in Fig. \ref{fig:fraction_target}.

Turning first to the 60 nm brushes, we observe that for all components and both brush shapes, we are able to consistently find designs which achieve the performance target. In most cases, only a few tens of optimization steps (and therefore simulations) are required. Random initialization affects the optimization trajectory, and the 20 different runs produce a distribution in the number of steps required to achieve the target. Comparing the circular and notched square brush, we find that with the circular brush the target performance is achieved slightly earlier, consistent with the hypothesis that the notched square presents a more challenging optimization problem. We observe this trend for other brush sizes discussed below. Across all component types designed with the 60 nm brush, the range of steps required to achieve the target performance is at most tens of steps wide, and in practical application a single optimization run may actually be sufficient to find a design which meets the desired performance specification.

When designing with a 80 nm brush, we again are able to consistently achieve the target performance. Generally, we observe that the number of required steps is larger than for the 60 nm brush. However, in some cases, such as the wavelength demultiplexer with circular brush, the number of steps required follow similar distribution as with the 60 nm brush. This suggests that for some problems, decreasing the length below a threshold value has no benefit. We also see some outlier examples where the target performance specification is not achieved.

As the brush sizes are increased to 100 nm, the general pattern is repeated. The target performance consistently achieved, but more steps are required and a larger number of outlier optimization runs fail to reach the target within 500 steps.
At a brush size of 130 nm, the trend is repeated yet again. With this brush size, designs for the waveguide bend and mode converter which achieve the target continue to be consistently found. For the beamsplitter and wavelength demultiplexer, however, approximately half or fewer optimization runs achieve the target. Therefore, in a practical application of our inverse design method with such large length scales, it may be beneficial to launch multiple optimization runs to ensure that at least one can reach the target performance. Additionally larger design regions may be necessary.

Much of the data supports the conclusion that larger brushes present more challenging optimization problems. We expect this also for theoretical reasons, since the design space with small brushes is strictly larger than that for larger brushes. However, with 130 nm brush we observe that some of the beamsplitter designs achieve the target performance ahead of the best 80 nm and 100 nm brush runs. This warrants further investigation in future work.

\section{Conclusion}
We have introduced an inverse design method which produces designs that are guaranteed to satisfy length scale constraints \textit{throughout} the course of an optimization. 
Our always-feasible method is based on a conditional generator for designs and a straight-through gradient estimator, which is commonly used in quantized neural network training \cite{bengio2013estimating}.
Combined, these two components enable the problem of length-scale-constrained topology optimization to be posed as an unconstrained stochastic gradient optimization problem.

While the computational graph of our optimization scheme (Fig. \ref{fig:graph}) resembles that of some established topology optimization methods \cite{Hammond2021-pv, Zhou2015-hg}, where a latent design is transformed into a physical design, the conditional generator used in our scheme avoids the challenges that established topology optimization techniques face in transitioning from performant \textit{infeasible} regions of the design space to performant \textit{feasible} designs.
In such high-dimensional design landscapes there is no guarantee that a high-performing infeasible design will be close to high-performing feasible designs, especially in integrated photonic devices where performance depends strongly on the collective interference of light from multiple scattering interfaces.
Density-based optimization schemes often employ an unconstrained optimization stage \cite{Hammond2021-pv, Vercruysse2019-ej} followed by multiple stages of increasing constraint enforcement, where device performance often drops significantly as design constraints are enabled.
While the optimizer is  often able to recover some level of performance, it is not guaranteed to be able to do so.
In some cases, the performance of the final feasible design remains lower than earlier infeasible designs.
A similar challenge in these optimization schemes lies in enforcing design binarization, which is often enabled gradually throughout several optimization stages by tuning a hyperparameter of the latent variable transformation \cite{Zhou2015-hg}.
While these changes to the device binarization reduce device performance, they present another challenge in that increased binarization is fundamentally at odds with the differentiability of the optimization graph. Essentially, the binarization nonlinearity approaches a step function with zero gradients.
The effect of this trade off is that during late stages of the optimization, the algorithm is not able to significantly change the topology of the structure.
In contrast, our always-feasible inverse design scheme always produces binarized designs and, moreover, is able to explore changes to the topology throughout the optimization (Fig. \ref{fig:bend}).

An added benefit of our always-feasible inverse design method is the user's ability to halt an optimization run once a design with sufficiently high performance is generated.
In contrast, obtaining a high performance design in a three field optimization scheme is typically \textit{not} a sufficient condition for halting an optimization because multiple rounds of infeasible optimization are required to achieve high performance.
Moreover, the higher levels of performance that are achieved during the unconstrained three field optimization stages are not guaranteed to be achieved during the stages where fabrication constraints and binarization have been applied.

The results presented in this paper demonstrate that our method can reliably find high-performing designs for a variety of 2D nanophotonic inverse design problems. We believe the good performance of our method, and its simplicity, could make it a useful new scheme in practical applications of inverse design. In the future, it would be important to use a realistic 3D electromagnetic simulator \cite{hughes2021perspective} to design components that can be manufactured in commercial foundries. Another interesting direction would be to apply the method to topology optimization problems in other physical domains.

We also see possible improvements to the underlying method. Specifically, conditional generators for designs satisfying additional constraints would be valuable, such as minimum solid area and void area \cite{Hammond2021-pv}, which are frequently included in foundry design rules. Even for width and spacing constraints addressed by the current scheme, it would be useful to develop new generators which are not conservative, i.e. which do not require a brush larger than the target minimum width. Finally, the estimator and transform functions warrant additional study, and we see potential for learned estimators.

\onecolumngrid
\clearpage

\section*{Supporting Information}

\renewcommand{\thesection}{S\arabic{section}}
\renewcommand{\thesubsection}{S\arabic{subsection}}
\setcounter{section}{0}
\renewcommand{\thefigure}{S\arabic{figure}}
\setcounter{figure}{0}
\renewcommand{\thetable}{S\arabic{table}} 
\setcounter{table}{0}
\renewcommand{\theequation}{S\arabic{equation}} 
\setcounter{equation}{0}
\renewcommand{\thepage}{S\arabic{page}} 
\setcounter{page}{1}

\section{Illustration of feasible design generation}
\label{appendix:traversal_example}

\begin{figure}[H]
    \centering
    \includegraphics{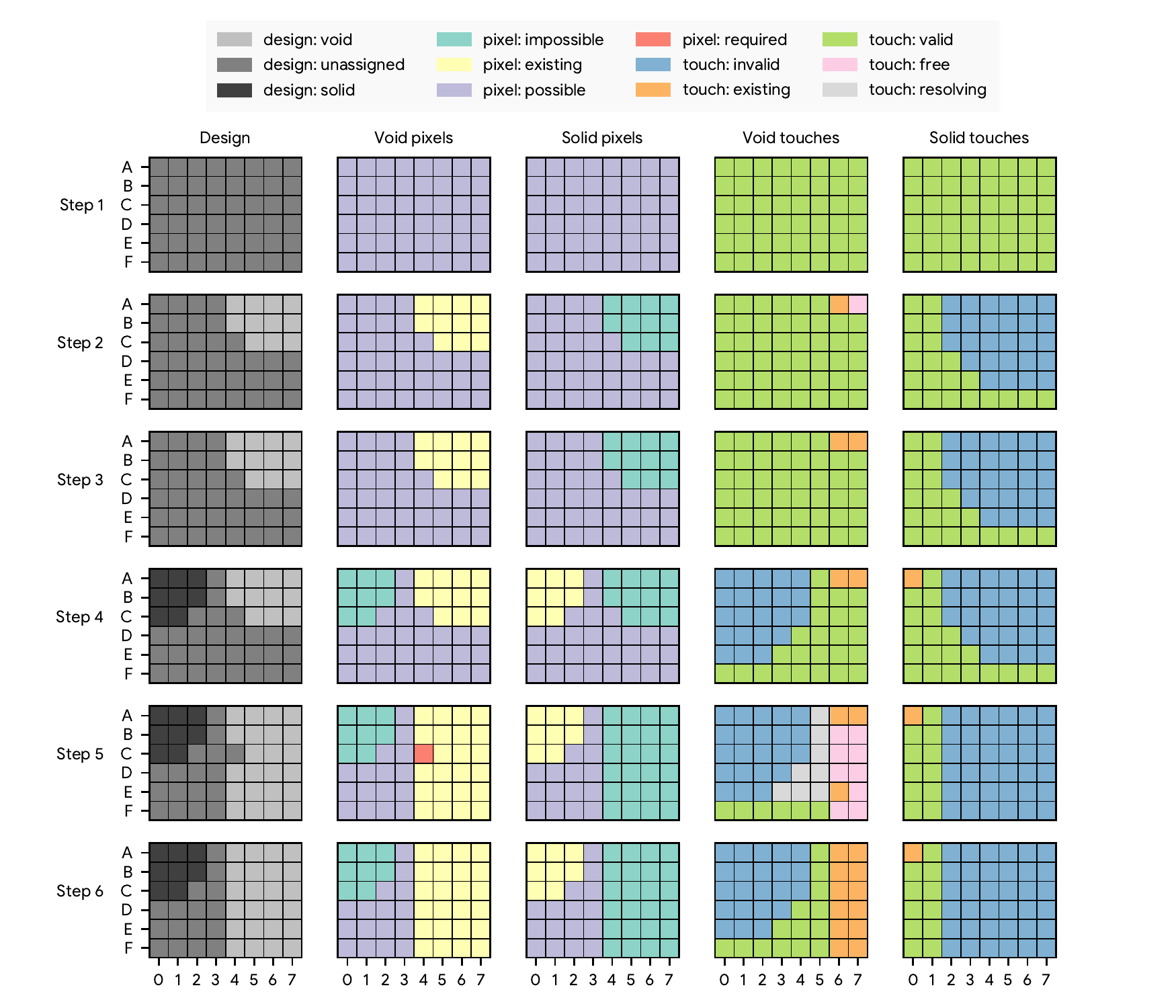}
    \caption{The first six steps of design generation. \emph{Step 1}: Solid and void touches are valid at all locations, and all pixels are possible. \emph{Step 2}: A void touch is placed at A6. Pixels covered by the touch become existing pixels for void, and impossible pixels for solid. A7 becomes a free touch for void. \emph{Step 3}: The free touch A7 is taken. The design and pixels are unaffected. \emph{Step 4}: A solid touch is placed at A0. \emph{Step 5}: A void touch is placed at E6. Location C4 becomes a required pixel for void. There are required-resolving touches in columns 3, 4, and 5, while there are free touches in columns 6 and 7. The resolving (but not free) touches would set C4 to void, but would also remove possible solid pixels. \emph{Step 6}: All the free touches are taken. The required pixel C4 is resolved, and no required-resolving touches remain.}
    \label{fig:traversal_1_to_6}
\end{figure}

\begin{figure}[H]
    \centering
    \includegraphics{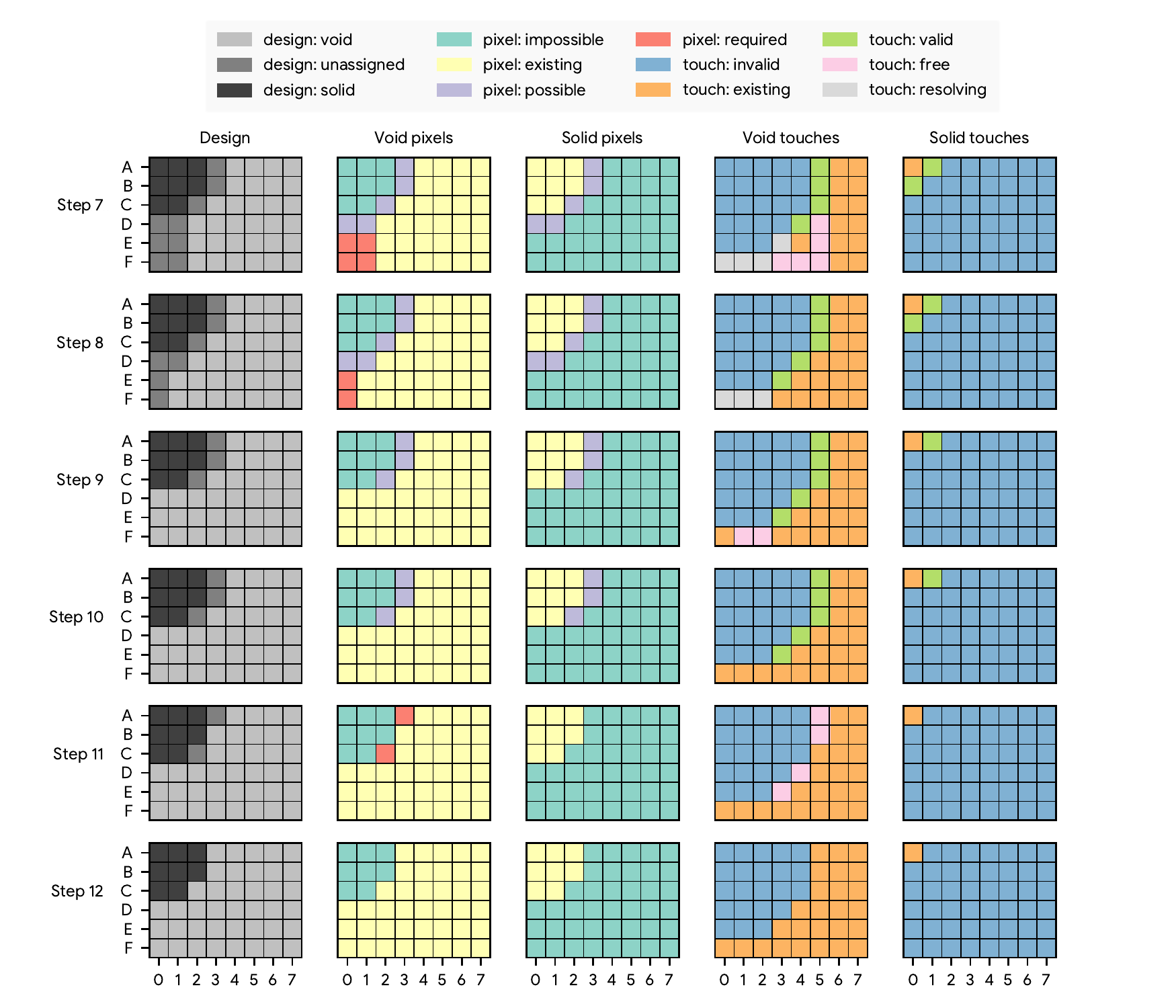}
    \caption{Additional steps of design generation. \emph{Step 7}: A void touch is placed at D4. Pixels E0, E1, F0, and F1 are now required for void. Both free and resolving (but not free) void touches exist. \emph{Step 8}: All the free touches are taken. E1 and F1 are resolved, but E0 and F0 remain required void pixels. \emph{Step 9}: A void touch is placed at F0, and the required pixels are resolved. F1 and F2 are now free void touches. \emph{Step 10}: The free touches at F1 and F2 are taken. \emph{Step 11}: A void touch is placed at C5. Pixels at A3 and C2 are now required for void. Since all remaining pixels must now be void, only free void touches remain. \emph{Step 12}: The free touches are made, and the design is complete.}
    \label{fig:traversal_7_to_12}
\end{figure}

\begin{figure}[H]
    \centering
    \includegraphics[width=0.6\columnwidth]{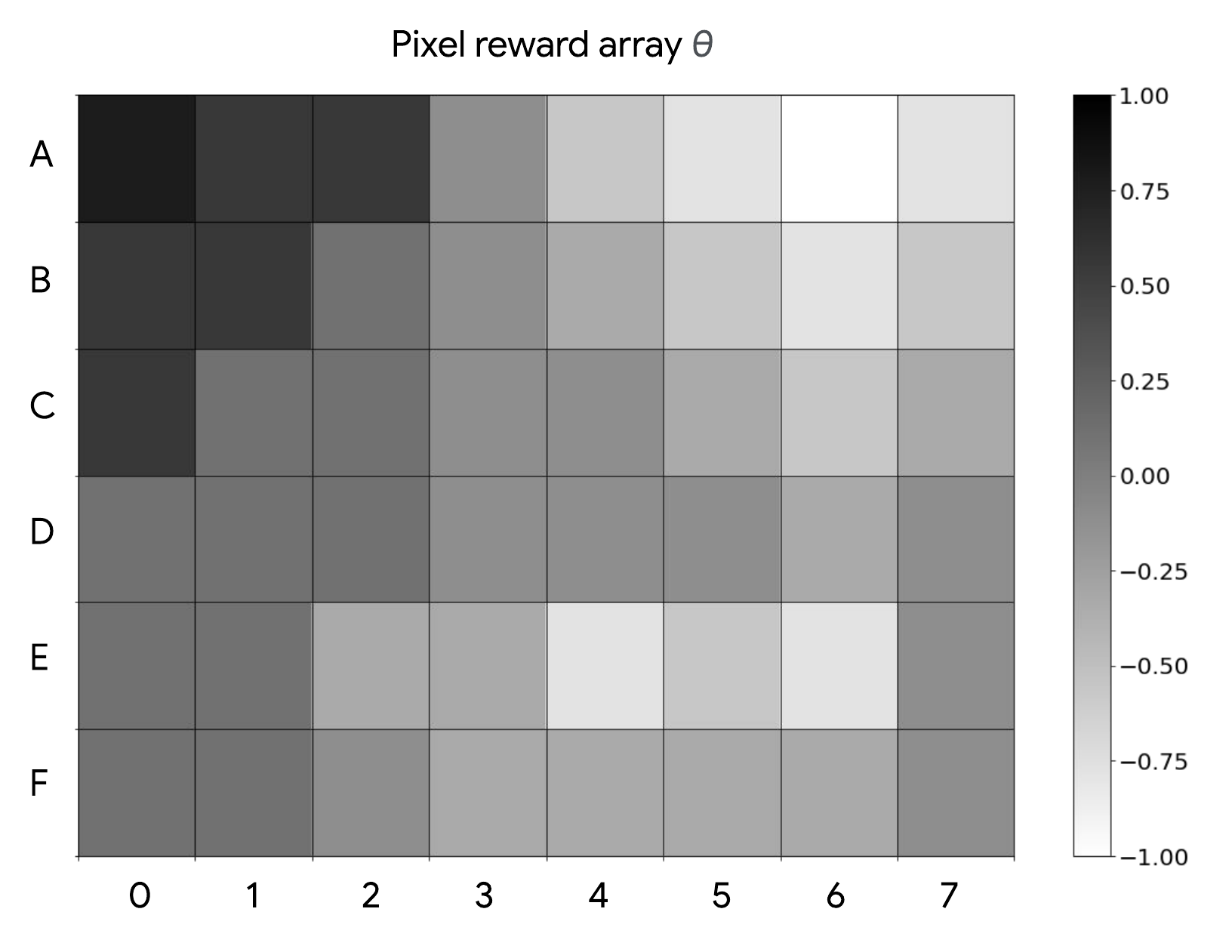}
    \caption{The continuous valued pixel reward array used to bias the 6$\times$8 feasible design generation stepped through in Figs. \ref{fig:traversal_1_to_6} and \ref{fig:traversal_7_to_12}}.
    \label{fig:theta}
\end{figure}

Figs. \ref{fig:traversal_1_to_6} and \ref{fig:traversal_7_to_12} illustrate the generation of a 6$\times$8 feasible design in 12 steps, using the algorithm from the main text conditioned on the input pixel reward array $\theta$ shown in Fig.~\ref{fig:theta}. A notched-square brush of width 5 pixels is used.

\clearpage

\clearpage

\section{Designs using 100 nm notched square brush}
\label{appendix:notched_square_brush}
\begin{figure}[H]
    \centering
    \includegraphics{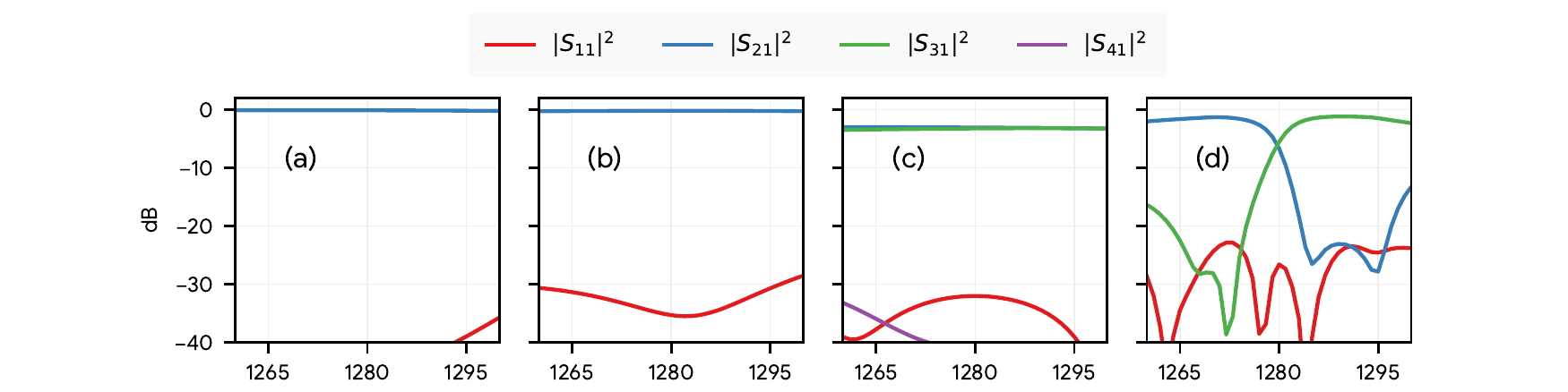}
    \caption{Scattering spectra for the (a) waveguide bend, (b) mode converter, (c) beamsplitter, and (d) wavelength demultiplexer with 100 nm notched square brush.}
    \label{fig:transmission_notched_square}
\end{figure}

\begin{figure}[H]
    \centering
    \includegraphics{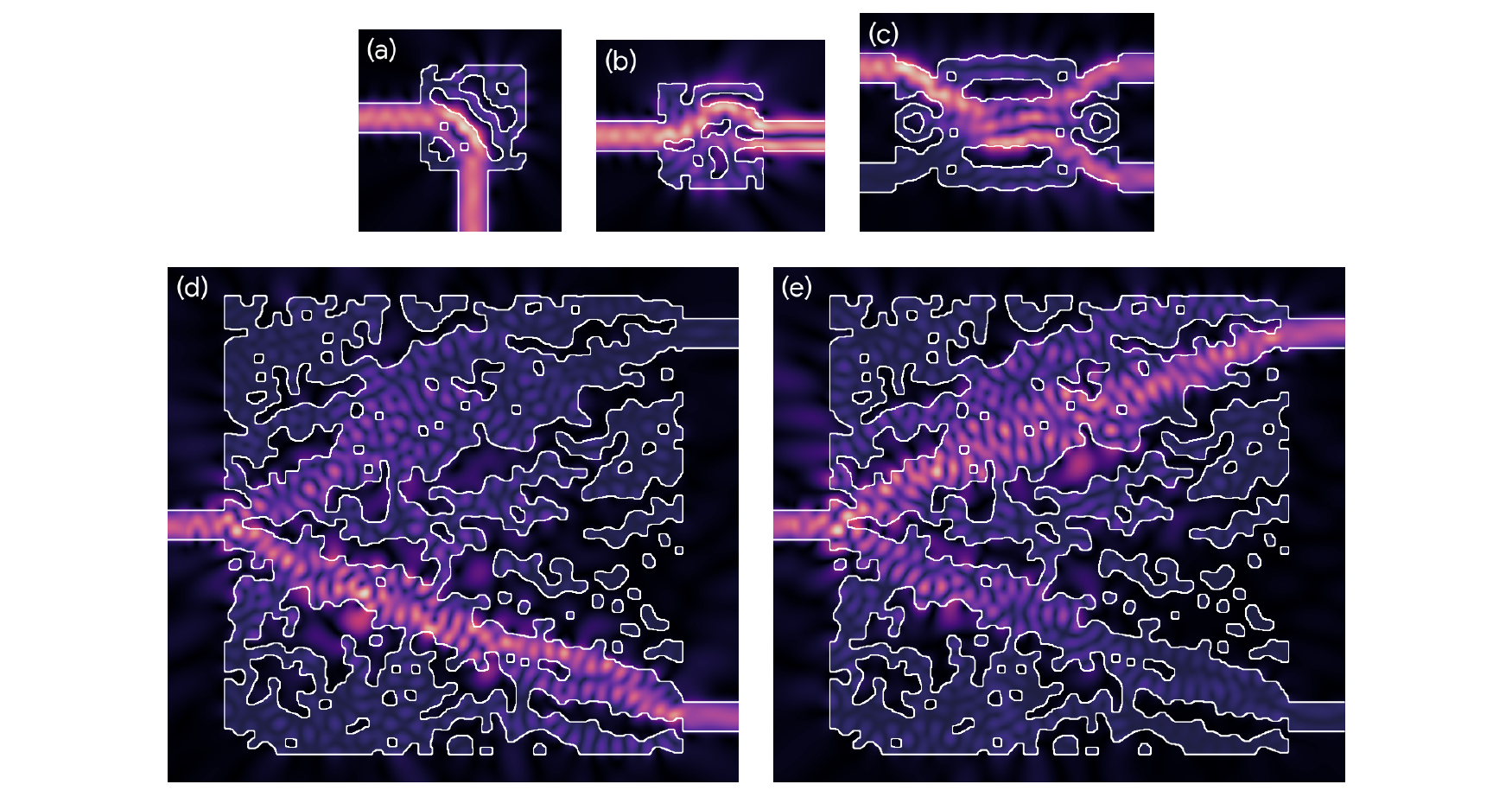}
    \caption{Lowest-loss designs achieved in the first 160 steps and their respective field magnitudes for (a) the waveguide bend, (b) mode converter, and (c) beamsplitter with 1280 nm excitation. Design and fields for the wavelength demultiplexer, with (d) 1270 nm and (e) 1290 nm excitation. All use a 100 nm notched square brush.}
    \label{fig:notched_square_designs}
\end{figure}

Scattering spectra for components created with a 100 nm notched square brush are shown in Fig. \ref{fig:transmission_notched_square}. The corresponding designs are shown in Fig. \ref{fig:notched_square_designs}. Each component is the lowest-loss component found in the first 160 steps of a randomly-initialized optimization run.

\section{Illustration of erosion and dilation}

\begin{figure}[H]
    \centering
    \includegraphics[width=4in]{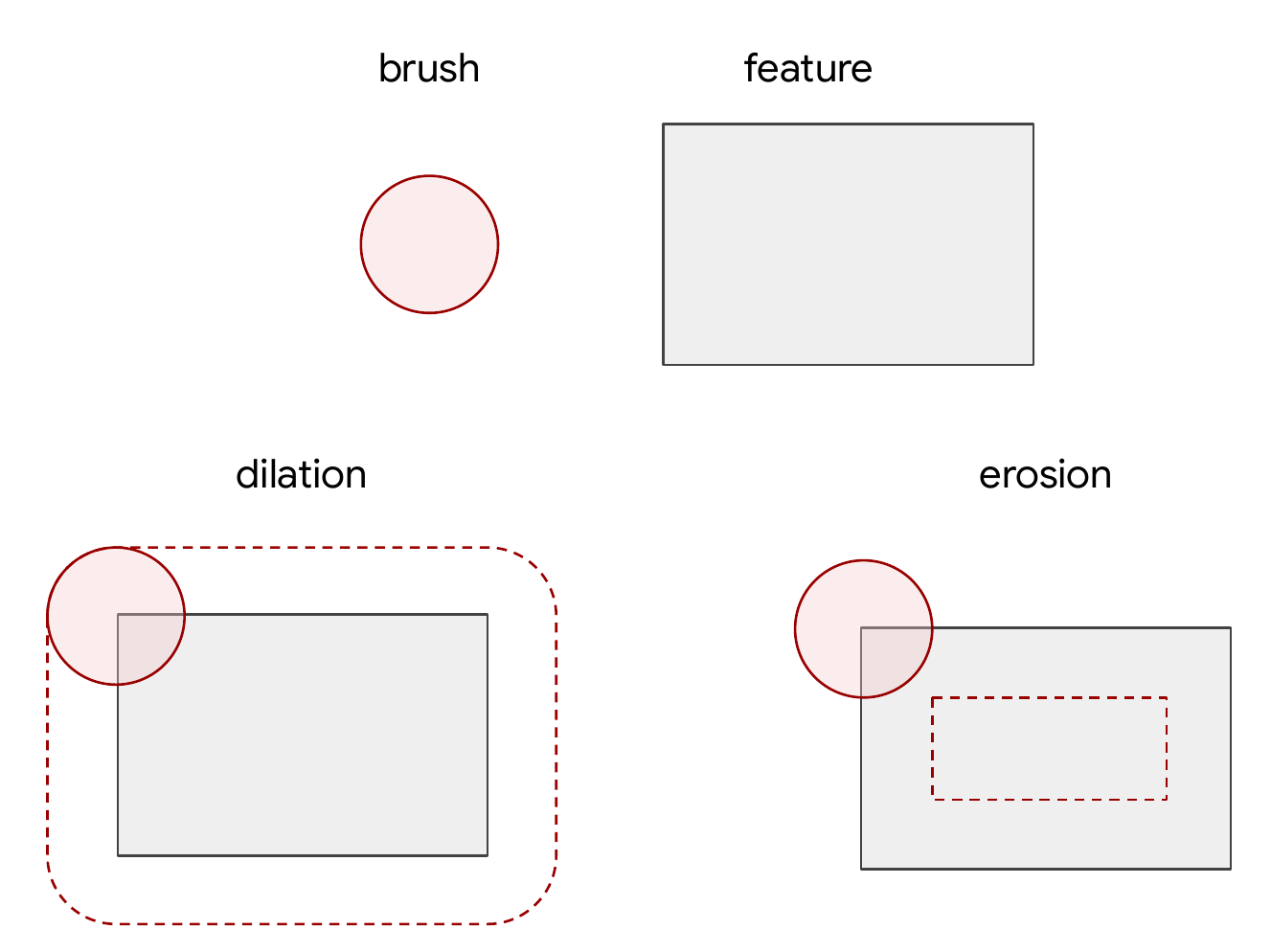}
    \caption{Illustration of erosion and dilation of a rectangular feature by a circular brush. The brush is translated around the boundary of the original feature to produce the dashed red outline, which corresponds to the eroded or dilated feature.}
    \label{fig:erosion_dilation}
\end{figure}

\clearpage

\section{Illustration of how rewards for touches are calculated}

\begin{figure}[H]
    \centering
    \includegraphics{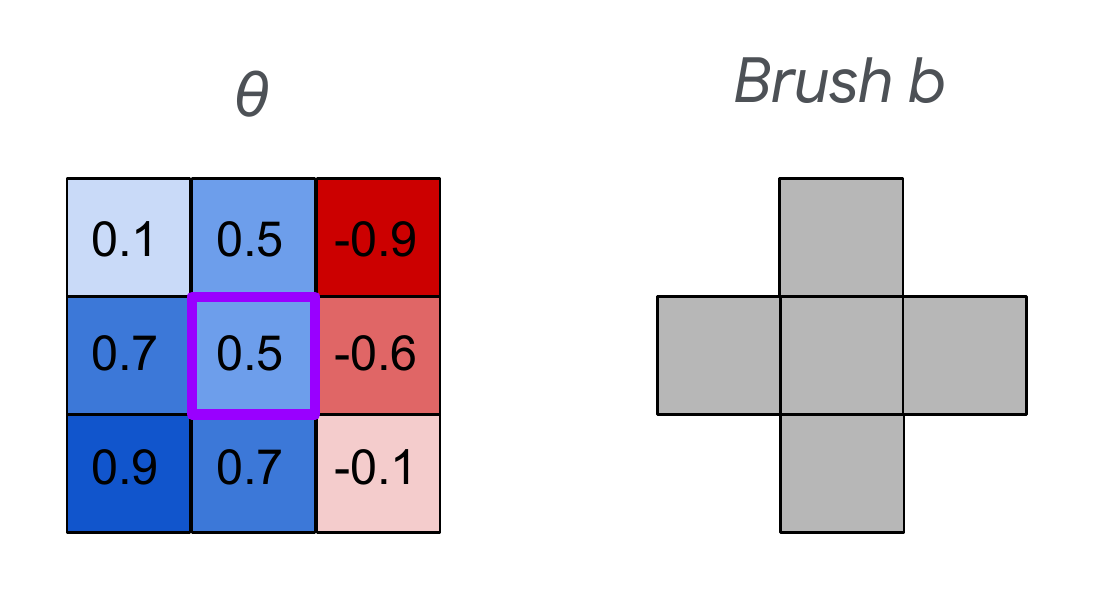}
    \caption{Simple example illustrating how rewards for otuches are calculated from the sum of $\theta$ elements set by the touch. In particular, the reward for placing a solid touch at the middle pixel (i.e. the pixel with the purple border) will be given by the sum of the sum of the values of $\theta$ over all the pixels covered by centering the brush at the middle pixel; i.e. it will be equal to 0.5 + 0.5 +0.7 + 0.7 + (-0.6) = 1.8. In contrast, the reward for placing a void touch at the middle pixel will be the negative sum (-1.8). Therefore, as the generator generates the feasible design, the middle pixel will more likely be set as a solid touch rather than a void touch, as expected from the predominantly positive valued (blue) $\theta$ in this example.}
    \label{fig:touch_cost}
\end{figure}

\end{document}